\documentclass{cta-author}

{}
{}
{}

\usepackage{amssymb}
\usepackage{setspace}
\usepackage{amsmath}
\usepackage[table]{xcolor}
\usepackage{colortbl}
\usepackage{color}
\usepackage{multirow}
\usepackage{multicol}
\usepackage{algorithm}
\usepackage{algpseudocode}
\usepackage{listings}
\usepackage[toc,page]{appendix}
\usepackage{empheq}
\usepackage{subfigure}
\usepackage{epstopdf}
\usepackage{graphicx}
\usepackage{url}

\usepackage{blkarray}

\usepackage{aliascnt}
\usepackage{remreset}

\usepackage{array}

\newtheorem{definition}{Definition}

\definecolor{listingsBackground}{rgb}{0.95,0.95,1.0} 

\lstdefinestyle{xmlCode}{language=XML,frame=tbrl,
  basicstyle=\tiny\ttfamily,keywordstyle=\tiny,
  stringstyle=\tiny\itshape,flexiblecolumns=false,
  commentstyle=\itshape,captionpos=b,abovecaptionskip=9pt,belowcaptionskip=0pt,
  backgroundcolor=\color{listingsBackground},frame=none, xleftmargin=6mm,
  numbers=left, numberstyle=\tiny}

\date{}

\begin{document}

\title{On-line tracing of XACML-based policy coverage criteria}

\author{\au{Francesca Lonetti}, \au{Eda Marchetti}}

\address{{Istituto di Scienza e Tecnologie dell'Informazione ``A.~Faedo'', CNR\\
  via G. Moruzzi 1, 56124 Pisa, Italy \\}
\email{firstname.secondname@isti.cnr.it}}

\begin{abstract}

Currently, XACML has becoming the standard for implementing access control policies and consequently more
attention is dedicated to testing the correctness of XACML policies.
In particular, coverage measures can be adopted for assessing test strategy effectiveness in exercising the policy elements.  This paper introduces a set of XACML coverage criteria and describes the  access control infrastructure, based on a monitor engine,  enabling the coverage criterion selection and the on-line tracing of the testing activity.
Examples of infrastructure usage and of assessment of different test strategies are provided.

\end{abstract}

\maketitle

\section{Introduction}

Security is a challenging issue in modern networked systems where adequate security mechanisms are put in place to preserve privacy and confidentiality of personal and critical data. Among security mechanisms, one
important component is the access control system, which mediates all requests of access to protected data.
Currently, among the various proposals, the OASIS eXtensible Access Control
Markup Language (XACML) \cite{xacml} is the most commonly used standard. However, the verbosity and complexity of XACML syntax
as well as the natural language semantics provided by the standard, make the specification and management of real access control policies in practice difficult and error-prone.

For preventing security flaws and violations, specific attention is devoted to the policy-based testing, i.e. the process to ensure the correctness of policy specifications and implementations. Indeed, by observing the execution of a policy implementation with a test input (i.e., an access request), the testers may identify faults in policy specifications or implementations, and validate whether the corresponding output (i.e., an access decision) is as intended.

Considering the strict constraints on testing budget, it is extremely important to focus the testing activity
on the generation or selection of the test cases that cover the most important features and/or policy constructs.
The purpose is to reduce as much as possible the number of tests to be executed
trying, from one side, to maximize the fault detection effectiveness, and from the other,
to cover the most important elements/aspects defined into the policy itself.
%

In this paper, we focus on the testing of the access control policies and in particular on the coverage assessment of the derived test suites.
In software testing, the coverage of specified entities is considered a valuable complement to the simple validation of the software behavior \cite{ammann2016introduction}. Indeed, the coverage information can provide an indication of the efficiency of the executed test cases and can help to
maintain an effective test suite.

In this paper, from one side, we define a set of coverage criteria that can help in the evaluation and assessment of the adopted testing strategy. On the other, we propose an access control policy infrastructure, based on an external monitoring facility, for enabling the language independent coverage measurement and automating the testing process itself.

A preliminary attempt to realize a coverage assessment framework through monitoring was presented in \cite{telerise17}.
Here we extend the result of \cite{telerise17} by defining four new coverage criteria for XACML policies and showing their application for the assessment of different testing strategies.
We present also the usage of the proposed infrastructure to realistic case studies.

The type of data to be collected during
the testing activity is independently specified by the execution engine and is not linked to the specific notation used for the policy specification.

The contribution of this paper can be summarized into: i) the definition of four original inclusive coverage criteria; ii) the integration of a monitoring framework into an access control system architecture; iii) the definition of the architecture of the Policy Assessment Infrastructure enabling the coverage criterion selection, the policy analysis, the monitoring of the policy execution, and the policy coverage assessment; iv) an instantiation of the proposed architecture on the XACML access control language.

The remainder of this paper is structured as follows: Section  \ref{basicConcepts} and \ref{coverageCriteria} introduce the basic concepts of access control systems and coverage criteria respectively;
Section \ref{framework} presents the architecture of the Policy Assessment Infrastructure;  Section \ref{example} presents experimental results on real XACML policies; finally, Section \ref{related} presents related work whereas Section \ref{conclusion} concludes the paper also hinting at future work.

\section{Basic Concepts}
\label{basicConcepts}

Access control is one of the most adopted security mechanisms for the protection of resources and data against unauthorized,
malicious or improper usage or modification. It is based on the implementation of access control policies expressed by a specific standard such for instance the widely adopted XACML~\cite{xacml}.

In this paper, we focus on the testing of the access control policies and in particular on the coverage
assessment of the derived test suite. Here below some basic concepts about the XACML-based access control system
and the coverage testing.

\begin{figure}[ht!]
 \centering
\includegraphics[width=\columnwidth]{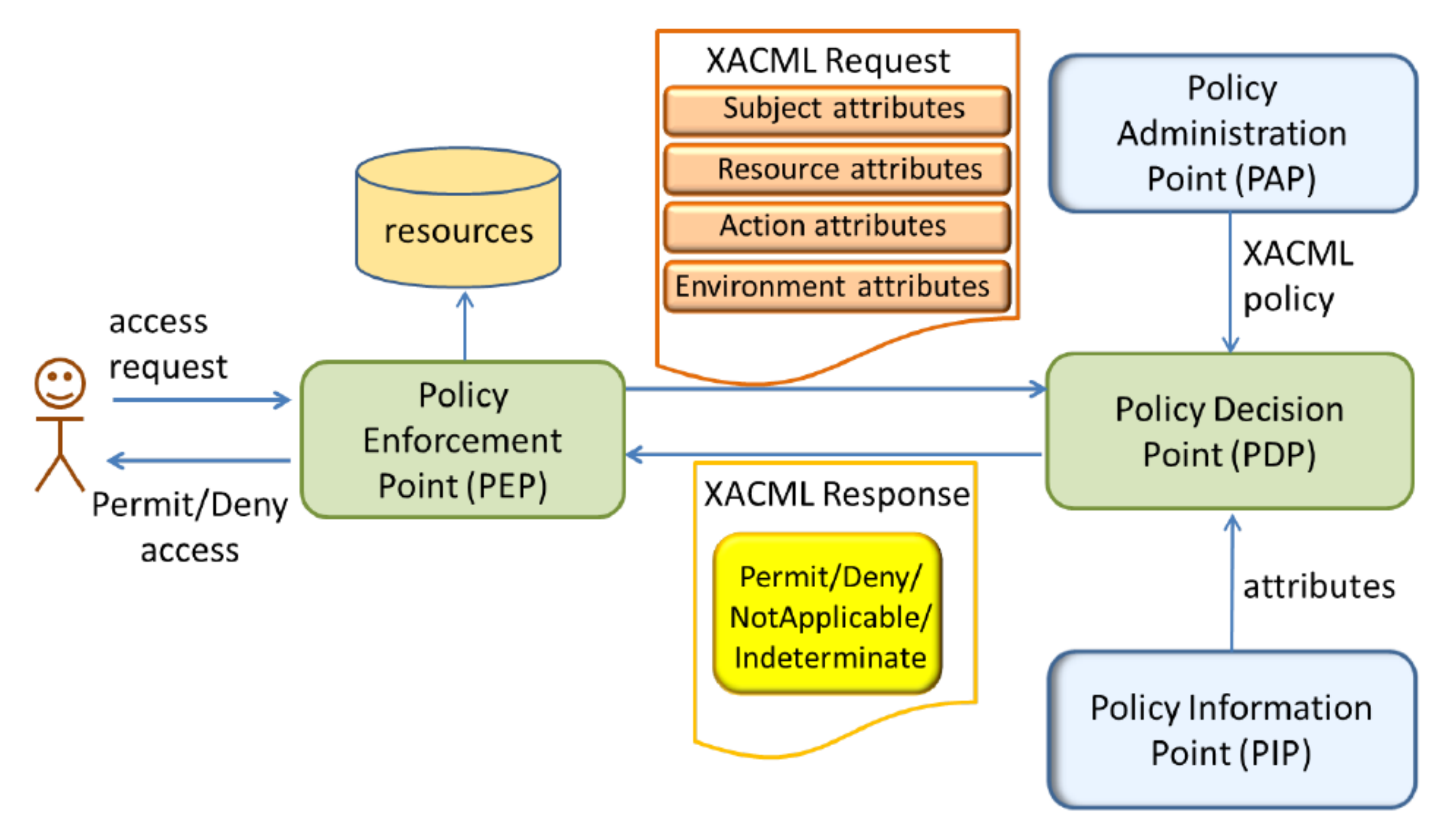}
 \caption{Access Control System Architecture}
 \label{architecture}
\end{figure}

XACML \cite{xacml} is a platform-independent XML-based language for the specification
of access control policies.
Briefly, an XACML policy has a tree structure whose main elements are: PolicySet, Policy, Rule, Target and Condition. The PolicySet includes one or more
policies. A Policy contains a Target and one or more rules. The Target specifies a
set of constraints on attributes of a given request. The Rule specifies a Target and a Condition containing one
or more boolean functions. If the Condition evaluates to true, then the Rule's
Effect (a value of \emph{Permit} or \emph{Deny}) is returned, otherwise a \emph{NotApplicable} decision
is formulated. The \emph{PolicyCombiningAlgorithm} and the \emph{RuleCombiningAlgorithm} define how to combine the results from multiple policies and rules respectively in order to derive a single access result.
An example of XACML policy is provided in Figure \ref{policy-example}. For more details see Section \ref{tworound}.
The main components of an XACML based access control system are shown in Figure~\ref{architecture}: the Policy
Administration Point (PAP) is the system entity in charge of managing the policies;
the Policy Enforcement Point (PEP) constructs an XACML
request and sends it to the Policy Decision Point (PDP); the Policy Information
Point (PIP) is in charge
of retrieving the attributes of users, resources, and environment  and sending them to the PDP;
the PDP evaluates the policy against the request and
returns the response, including the authorization decision to the PEP.
For more details we refer to XACML 2.0 standard \cite{xacml}.

Measurement of test quality is one of the key issues in software testing and coverage measures represent an effective mean for evaluating the different testing approaches \cite{ammann2016introduction}. Adequacy criteria evaluate the testing
strategy through the percentage of exercised set of elements in
the program or in the specification.
Usually, test coverage can be used for different purposes:
i) improve the test suite so to exercise  elements that have not been tested;
ii) test suite augmentation and test suite minimization in case of regression testing;
iii) test cases selection, prioritization and test suite effectiveness evaluation.
A systematic review of coverage based testing is presented in \cite{shahid2011study}.

\section{Coverage Criteria}
\label{coverageCriteria}

\begin{figure*}[t!]
  \centering
  \includegraphics[width=0.85\textwidth]{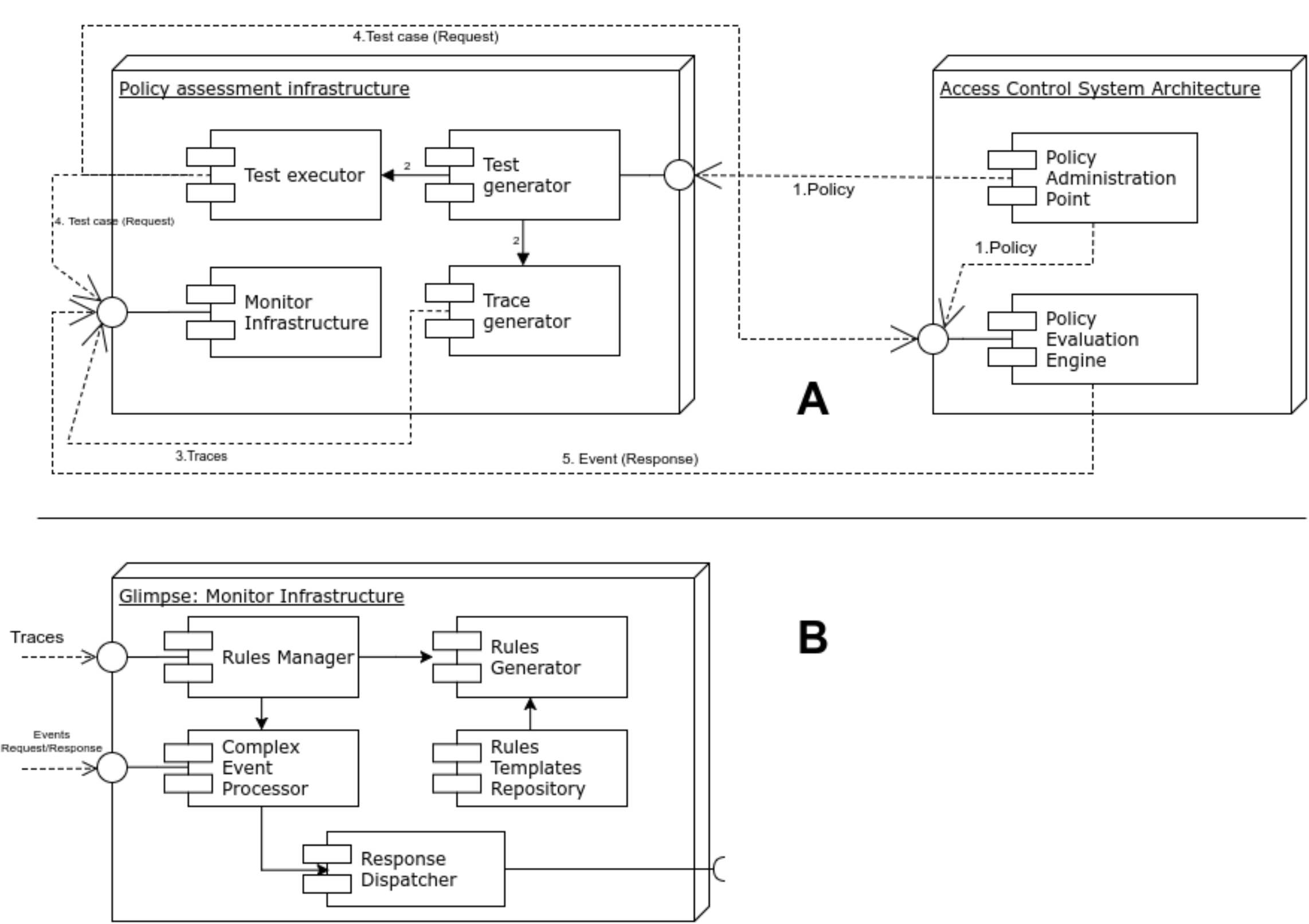}
    \caption{Policy Assessment Infrastructure}
  \label{fig:msm}
\end{figure*}

This section provides a set of XACML coverage criteria useful for assessing the effectiveness of a generic XACML based testing strategy.
Revising and extending the definition provided in \cite{bertolino2014coverage}, we first provide some generic definitions concerning the policy (Definitions \ref{def:target} and \ref{def:targetSet}) and request elements (Definition \ref{def:requestTarget}) and then we define the XACML coverage criteria (Definitions \ref{def:rtt}, \ref{def:rcont}, \ref{def:rtf}, \ref{def:rcf}).

\begin{definition}[Target Tuple]
\label{def:target}
Given a Rule \emph{R}, a Policy \emph{P}, a PolicySet \emph{PS}, with \emph{R} $\in$ \emph{P} and \emph{P} $\in$ \emph{PS}, and given the set of XACML Elements, called XE = \{xe : xe is \emph{PS} or \emph{P} or \emph{R}\},  the Target Tuple of an xe $\in$ XE, called $TT_{xe}$, is a 4-tuple ($S$, $Res$, $A$, $E$), where:
$S$ ($Res$, $A$, $E$) is a finite set of subjects (resources, actions, environments) in the XACML target of xe.
\end{definition}

\begin{figure}
\includegraphics[width=0.97\columnwidth]{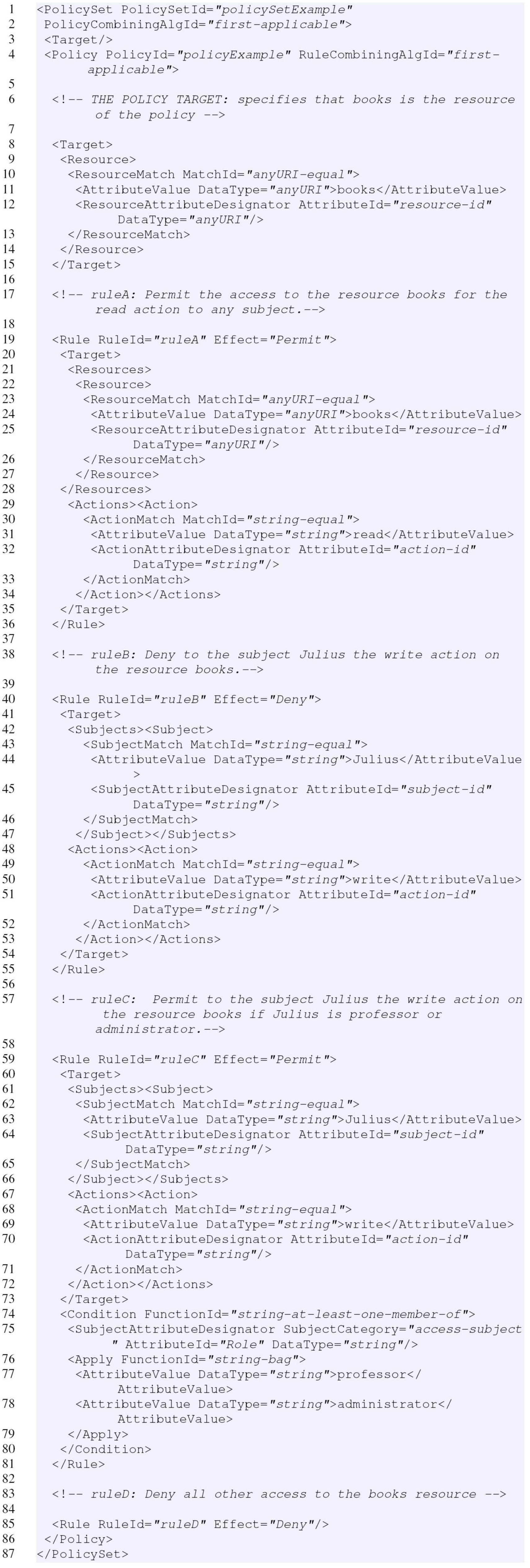}
\caption{An XACML Policy}
\label{policy-example}
\end{figure}

For instance, considering the policy of Figure \ref{policy-example}, the Target Tuple of the policy \emph{P} is equal to ($\emptyset$, \{$books$\}, $\emptyset$, $\emptyset$).

\begin{definition}[Rule Target Set]
\label{def:targetSet}
Given a Rule \emph{R}, its Target Set is a set of Target Tuple,
ordered by the XACML hierarchy elements relation, defined as
\begin{displaymath}  \;   TS_{R}=
 \left\{
   \begin{array}{lll}
        TT_{xe}:& & TT_{xe} = \left\{
	    \begin{array}{lll}
		TT_{PS} &\; & \text{if} \; R \in PS \\
		TT_{P}  &\; & \text{if} \; R \in P \\
		TT_{R} &\; & \text{otherwise } \;  \\
	    \end{array}
	    \right.
    \end{array}
\right\}.
\end{displaymath}
\end{definition}

For instance, considering the policy of Figure \ref{policy-example}, the Rule Target Set of \emph{RuleA} is equal to:

\{($\emptyset$,$\emptyset$,$\emptyset$,$\emptyset$), ($\emptyset$,\{$books$\},$\emptyset$,$\emptyset$), ($\emptyset$,\{$books$\},\{$read$\},$\emptyset$)\}

whereas the Rule Target Set of \emph{RuleB} is equal to:

\{($\emptyset$,$\emptyset$,$\emptyset$,$\emptyset$), ($\emptyset$,\{$books$\},$\emptyset$,$\emptyset$), (\{$Julius$\},\{$books$\},\{$write$\},$\emptyset$)\}.

\begin{definition}[Rule Effect]
\label{targeteffectrule}
Given a Rule \emph{R}, and given a set RE = $\{$ re : re is \emph{Permit} or \emph{Deny}$\}$,  the Rule Effect of  \emph{R}, called $RE_{R}$, is the re $\in$ RE, such that re is equal to the \emph{Effect}  of  \emph{R}.
\end{definition}

For instance, considering the policy of Figure \ref{policy-example}, the Rule Effect of \emph{RuleA} is equal to \emph{Permit} (see line 19).

\begin{definition}[Request Target Tuple]
\label{def:requestTarget}
Given a request $Req$, the Request Target Tuple, called $TT_{req}$ is a 4 tuple ($S\_r$, $Res\_r$, $A\_r$, $E\_r$) where $S\_r$, $Res\_r$, $A\_r$, $E\_r$ are the sets of subjects, resources, actions and environments belonging to the request $Req$.
\end{definition}

\begin{figure}
\includegraphics[width=\columnwidth]{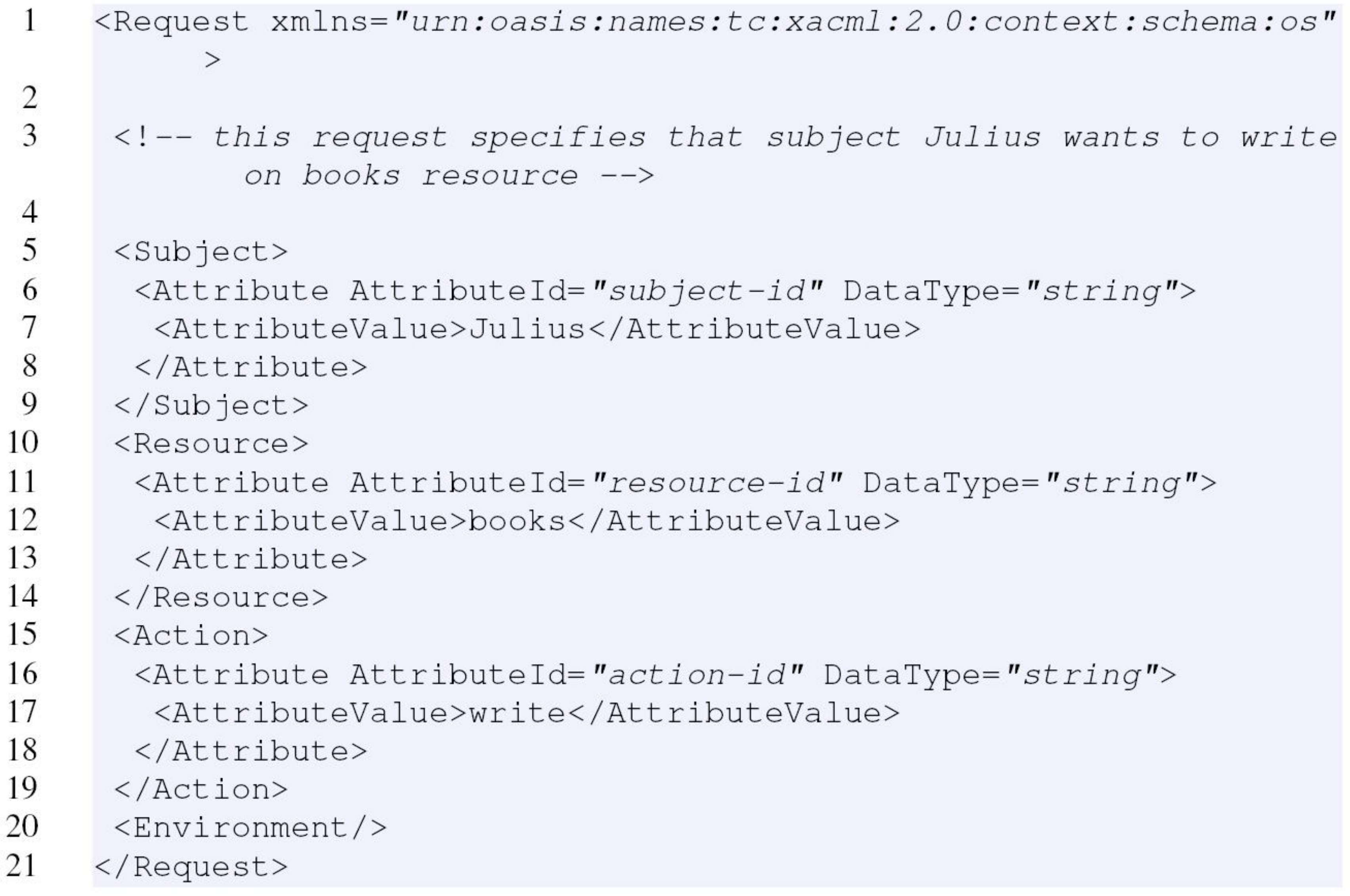}
\caption{An XACML request generated by \emph{Simple Combinatorial}}
\label{request-exampleSimple}
\end{figure}

For instance, considering the request of Figure \ref{request-exampleSimple}, the Request Target Tuple is equal to (\{$Julius$\}, \{$books$\}, \{$write$\}, $\emptyset$).

Using the above definitions in the following the different  XACML rule coverage criteria are introduced as well as the definition of their effects.

In a nutshell, the  XACML rule coverage criteria involve selecting tests that match or not the \emph{Rule Target Set}s. The Rule Target Set is the union of the target of the rule, and all enclosing policy and policy sets targets. The main idea is that according to the XACML language in order to match the rule target, requests must first match the enclosing policy and policy sets targets.

\begin{definition}[Rule Target True Criterion]
\label{def:rtt}

Given a rule $R$, the  Rule Target Set $TS_{R}$, and the request $Req$ with Request Target Tuple $TT_{req}$=($S\_r$, $Res\_r$, $A\_r$, $E\_r$), $Req$ covers the $TS_{R}$ with value true if and only if for each Target Tuple $TT_{E}$ =($S$, $Res$, $A$, $E$) $\in$ $TS_{R}$ such that  $TT_{E}$  is a $TT_{PS}$, $TT_{P}$ or $TT_{R}$
\begin{itemize}
\item $\exists$ an element s $\in$ S\_r   such that s $\in$ $S$ or $S$ is $\emptyset$, and
\item $\exists$ an element r $\in$ Res\_r such that r $\in$ $Res$ or $Res$ is $\emptyset$, and
\item $\exists$ an element a $\in$ A\_r    such that a  $\in$ $A$ or $A$ is $\emptyset$, and
\item $\exists$ an element e $\in$ E\_r  such that e   $\in$ $E$ or $E$ is $\emptyset$.
\end{itemize}

\end{definition}

For instance, considering the policy of Figure \ref{policy-example}, the Rule Target Set of \emph{RuleB}  (see above) and the request of Figure \ref{request-exampleSimple}, that request satisfies the Rule Target True Criterion of \emph{RuleB}. This request does not satisfies the Rule Target True Criterion of \emph{RuleA} because the \emph{write} action of the request does not belong to the set of actions of Rule Target Set of \emph{RuleA}.

\begin{definition}[Rule Target True Effect]
\label{def:rtte}
Given a rule $R$,  the condition $C$ of $R$, the  Rule Target Set $TS_{R}$, the Rule Effect $RE_{R}$, and the request $Req$ with Request Target Tuple $TT_{req}$, such that $Req$ satisfies the Rule Target True criterion, the Rule Target True Effect is
\begin{itemize}
  \item $RE_{R}$ if $C$ is evaluated to True against $TT_{req}$
  \item $\emptyset$ otherwise
  \end{itemize}
\end{definition}

If the rule has no condition or the condition can be evaluated to true, the Rule Target True Effect is equal to the effect of the rule.

\begin{definition}[Rule Condition True Criterion]
\label{def:rcont}

Given a rule $R$,  the condition $C$ of $R$, the  Rule Target Set $TS_{R}$, and the request $Req$   with Request Target Tuple $TT_{req}$, such that $Req$ satisfies the Rule Target True criterion,  the request  $Req$ covers the $C$ with value true if and only if $C$ is evaluated to True against $TT_{req}$.
\end{definition}

\begin{definition}[Rule Condition True Effect]
\label{def:rct}
Given a rule $R$,  the condition $C$ of $R$, the  Rule Target Set $TS_{R}$, the Rule Effect $RE_{R}$, and the request $Req$  with Request Target Tuple $TT_{req}$ the Rule Condition True Effect is  $RE_{R}$ if and only if:
\begin{itemize}
\item $Req$ satisfies the Rule Target True criterion and,
\item $C$ is evaluated to True against $TT_{req}$
 \end{itemize}

\end{definition}

\begin{figure}
\includegraphics[width=\columnwidth]{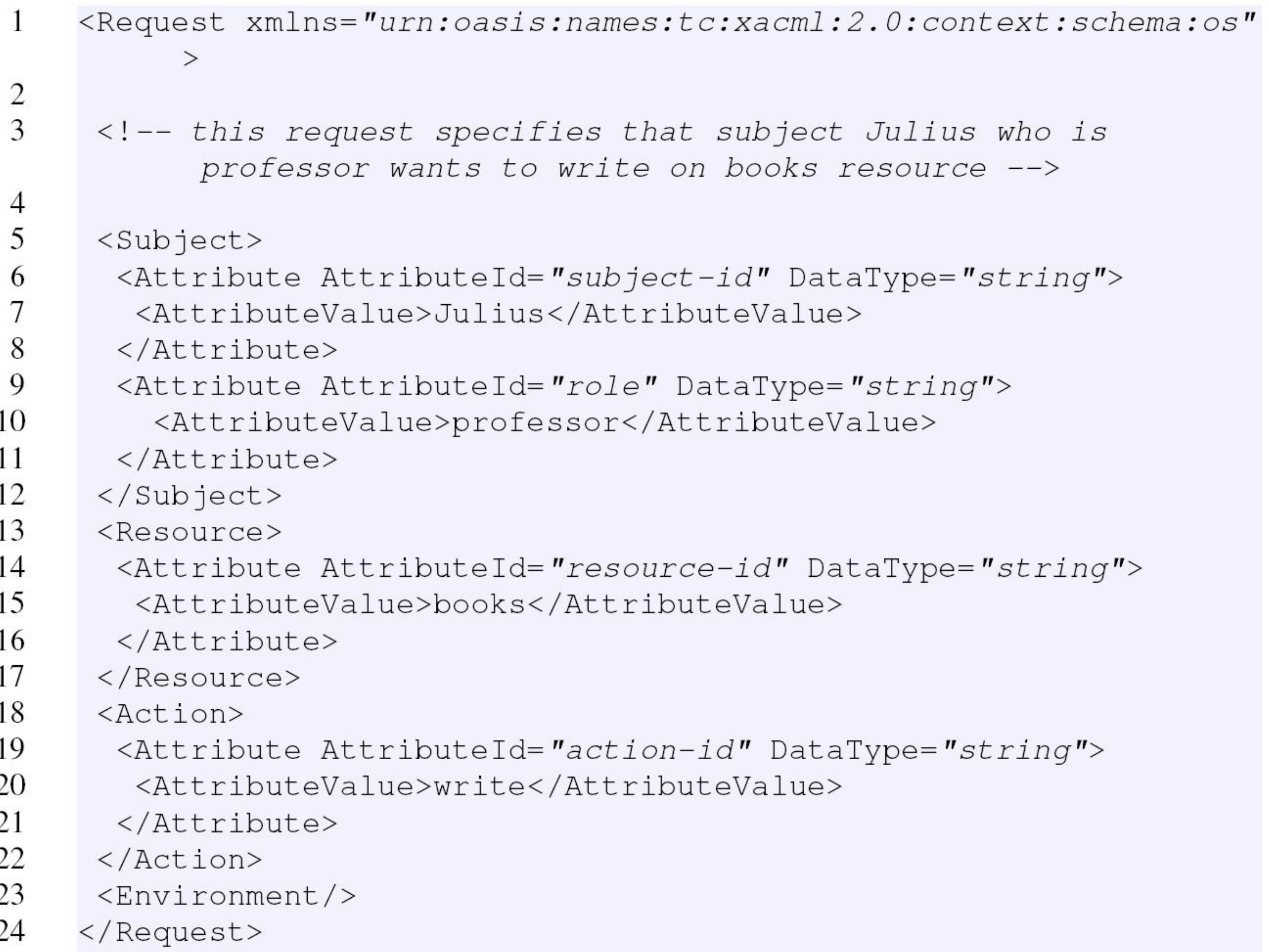}
\caption{An XACML request generated by \emph{Multiple Combinatorial}}
\label{request-exampleMultiple}
\end{figure}

Referring to policy of Figure \ref{policy-example}, and the \emph{RuleC}, the request of Listings \ref{request-exampleMultiple} having Request Target Tuple equal to (\{$Julius$ $\land$ $professor$\}, \{$books$\}, \{$write$\}, $\emptyset$) covers the Rule Condition True criterion.

In this case, the Rule Condition True Effect of \emph{RuleC} is \emph{Permit} (line 59).

\begin{definition}[Rule Target False Criterion]
\label{def:rtf}

Given a rule $R$,  the  Rule Target Set $TS_{R}$, and the request $Req$ with Request Target Tuple $TT_{req}$=($S\_r$, $Res\_r$, $A\_r$, $E\_r$), the $Req$ covers the $TS_{R}$ with value false if and only if for each Target Tuple $TT_{E}$ =($S$, $Res$, $A$, $E$) $\in$ $TS_{R}$ such that  $TT_{E}$  is a $TT_{PS}$ or $TT_{P}$

\begin{itemize}
\item $\exists$ an element s $\in$ S\_r   such that s $\in$ $S$ or $S$ is $\emptyset$, and
\item $\exists$ an element r $\in$ Res\_r such that r $\in$ $Res$ or $Res$ is $\emptyset$, and
\item $\exists$ an element a $\in$ A\_r    such that a  $\in$ $A$ or $A$ is $\emptyset$, and
\item $\exists$ an element e $\in$ E\_r  such that e   $\in$ $E$ or $E$ is $\emptyset$.
\end{itemize}

and for  $TT_{E}$  equal to $TT_{R}$,

\begin{itemize}
\item $\nexists$ an element s $\in$ S\_r   such that s $\in$ $S$, or
\item $\nexists$ an element r $\in$ Res\_r such that r $\in$ $Res$ or
\item $\nexists$ an element a $\in$ A\_r    such that a  $\in$ $A$ or
\item $\nexists$ an element e $\in$ E\_r  such that e   $\in$ $E$.
\end{itemize}

\end{definition}

For instance, considering the policy of Figure \ref{policy-example}, the Rule Target Set of \emph{RuleA}  and the request of Figure \ref{request-exampleSimple}, that request satisfies the Rule Target False Criterion of \emph{RuleA} because the \emph{write} action of the request does not belong to the set of actions of Rule Target Set of \emph{RuleA}.

\begin{definition}[Rule Target False Effect]
\label{def:rtfe}
Given a rule $R$,  the  Rule Target Set $TS_{R}$, the Rule Effect $RE_{R}$, and the request $Req$   such that $Req$ satisfies the Rule Target False, the Rule Target False Effect is always $\emptyset$.
\end{definition}

\begin{definition}[Rule Condition False Criterion]
\label{def:rcf}

Given a rule $R$,  the condition $C$ of $R$, the  Rule Target Set $TS_{R}$, and the request $Req$ with Request Target Tuple $TT_{req}$  such that $Req$  satisfies the Rule Target True criterion,  the request  $Req$ covers the $C$ with value false if and only if $C$ is evaluated to false against $TT_{req}$.
\end{definition}

For instance, considering the policy of Figure \ref{policy-example}, and the \emph{RuleC}, the request of Figure \ref{request-exampleSimple} satisfies the Rule Condition False Criterion of \emph{RuleC} because: the request satisfies  the Rule Target True criterion for  \emph{RuleC} but the sets of subjects of the request do not include \emph{professor} or \emph{administrator} values.

\begin{definition}[Rule Condition False Effect]
\label{def:rcfe}
Given a rule $R$, the condition $C$ of $R$, the  Rule Target Set $TS_{R}$, the Rule Effect $RE_{R}$, and the request $Req$   such that $Req$ satisfies the Rule Condition False, the Rule Target False Effect is always $\emptyset$.
\end{definition}

\section{Coverage Testing Framework}
\label{framework}

In this section, we present the architecture of Policy Assessment Infrastructure based on an on-line monitor.
The proposal has been conceived to be independent from the language adopted for the policy specification and flexible enough to be adapted to the different testing purposes.
In particular, Figure \ref{fig:msm} (part A) shows the main components of the proposed Policy Assessment Infrastructure (top left component), referring to a generic structure of an access control system (top right component):
\begin{itemize}
\item \emph{Test case generator} is in charge of  test cases generation starting from the policy specification. In literature, depending on the access policy language there are several proposals such for instance XCREATE \cite{bertolino2012automatic} and Targen \cite{Martin:2006:ATG:1176617.1176708} focused on XACML-based combinatorial approaches.
\item \emph{Test case executor} takes in input the test suite derived by the \emph{Test case generator}, and sends one by one the test cases to the \emph{Policy executor engine}. Moreover, it extracts the required information by each test case and transforms them into events readable by the \emph{Monitor infrastructure}.

\item \emph{Trace generator}  is in charge of deriving the set of policy traces. Specifically,
it takes in input the policy from the \emph{Policy Administration Point} and, according to the selected
coverage criterion, derives all the possible policy traces. Each trace is a target tuple able to satisfy the selected criterion plus the associated resulting criterion effect according to the definitions of  Section \ref{coverageCriteria}.
    Usually, the traces extraction is realized by an optimized unfolding algorithm
 that exploits  the policy language structure. Intuitively, the main goal is to derive an acyclic graph, defining a partial order on policy elements. Several proposals are available such as \cite{martin2006defining,bertolino2014coverage} for XACML policy specification. Once extracted, the traces are provided to the \emph{Monitor infrastructure}.

\item \emph{Policy evaluation engine} is in charge of the execution of the policy and the derivation of the associated response. It communicates with the \emph{Monitoring infrastructure} though a dedicated interface such as a REST  one.

\item \emph{Monitoring infrastructure} is in charge of collecting data of interest during the run-time policy execution.
 There can be different solutions for monitoring activity.  In this paper, we rely on Glimpse \cite{bertolino2011towards} monitoring infrastructure which
  has the peculiarity of decoupling the events specification from their collection and processing. As detailed in Figure \ref{fig:msm} (part B), the main components of Glimpse are: i) \emph{Complex Events Processor (CEP)} which analyzes the events and correlates them to infer more complex events; ii) \emph{Response Dispatcher}  keeps track of the monitoring requests and sends back the final coverage evaluation; iii) \emph{Rules Generator} generates the rules using the templates stored into the \emph{Rules Template Repository} starting from the derived policy traces to be monitored. A generic rule consists of two main parts: the events to be matched and the constraints to be verified, and the events/actions to be notified after the rule evaluation; iv) \emph{Rules Template Repository} stores predetermined rules templates that will be instantiated by the \emph{Rules Generator} when needed; v) \emph{Rules Manager}  instructs the CEP by loading and unloading the set of rules. We refer to \cite{bertolino2011towards} for a more detailed description of the Glimpse architecture.

 \end{itemize}

In a typical workflow  of the proposed framework, the \emph{Policy Administration Point} sends the policy both to the \emph{Test case generator} and the \emph{Policy evaluation engine} (step 1 of Figure \ref{fig:msm}). The \emph{Test case generator} derives from the policy specification a set of test cases and sends them to both the \emph{Test case executor} and the \emph{Trace generator} (step 2 of Figure \ref{fig:msm}). The \emph{Trace generator} derives from the policy all the possible policy traces and sends them to the \emph{Monitoring infrastructure} (step 3 \emph{Monitoring infrastructure}). The \emph{Test case executor} sends the test cases to the \emph{Policy evaluation engine}, moreover it extracts from these test cases the events that are forwarded to the \emph{Monitoring infrastructure} (step 4 of Figure \ref{fig:msm}). Finally, the responses associated to the execution of the test cases are forwarded by the \emph{Policy evaluation engine} to the \emph{Monitoring infrastructure} (step 5 of Figure \ref{fig:msm}).

\section{Experimental results}
\label{example}

%

\begin{figure}
\includegraphics[width=\columnwidth]{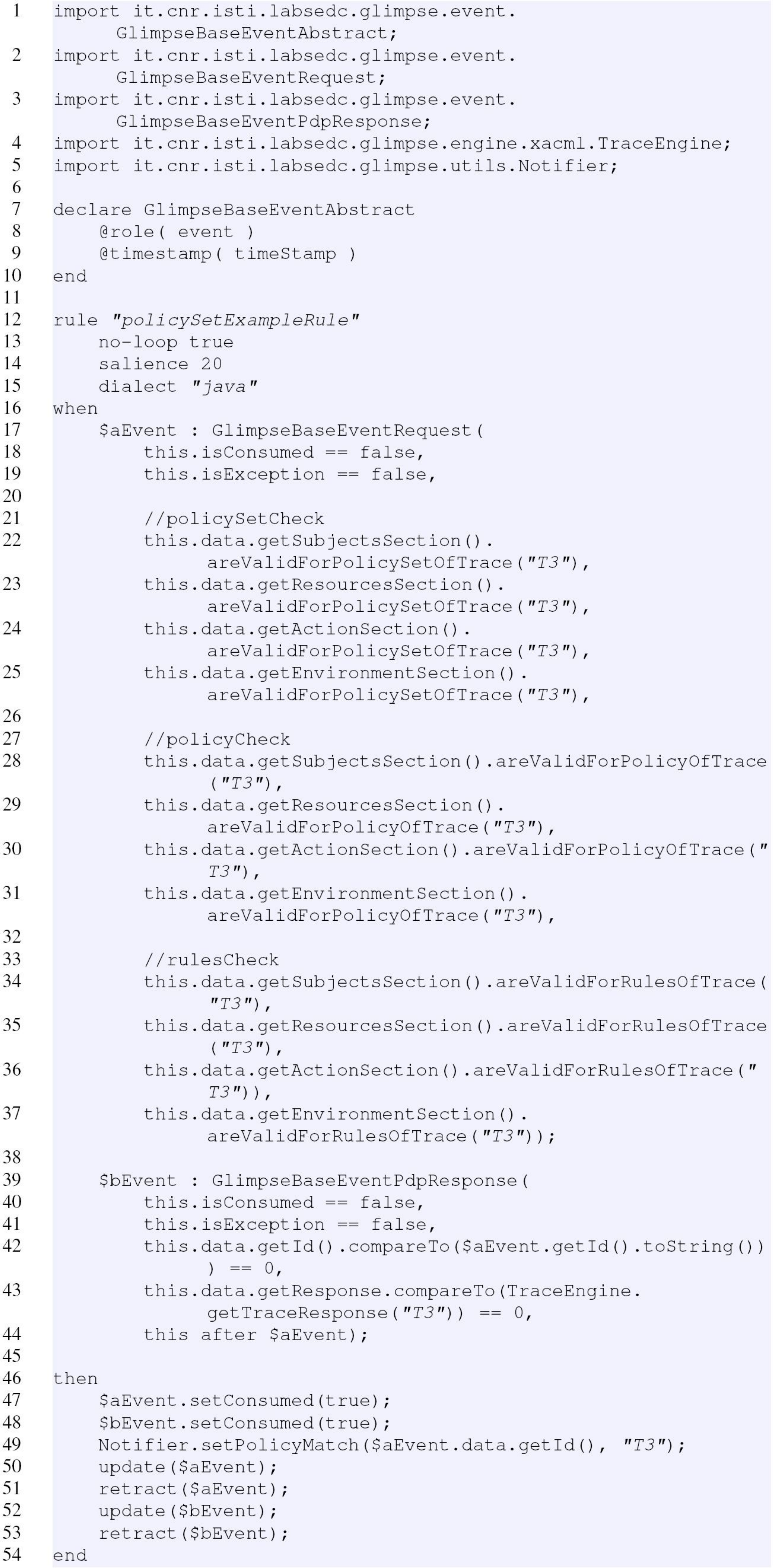}
\caption{Monitoring Rule}
\label{rule}
\end{figure}

In this section, the Policy Assessment Infrastructure, instantiated on the XACML based access control systems, has been used for evaluating the coverage measures of two different test strategies (Section \ref{strategies}). In particular, we first present a two-round application example so to better describe the infrastructure behavior and features (Section \ref{tworound}). Then the coverage evaluation of the test strategies with real word access control policies is presented (Section \ref{experiment}).

\subsection{Testing strategies}
\label{strategies}

Several common approaches for generating XACML requests are based on combinatorial strategies \cite{iet}. In this paper, among the tools available for test cases generation we refer to  X-CREATE~\cite{seaa10,bertolino2012automatic,iet}.
 In particular, we use two test strategies implemented in this tool that are \emph{Simple Combinatorial} and \emph{Multiple Combinatorial}, for deriving the test suites used to empirically validate the effectiveness of the proposed approach.
We provide below a brief description of the two strategies referring to \cite{iet} for more details.

\subsubsection{Simple Combinatorial}

The \emph{Simple combinatorial} strategy applies a combinatorial approach to the policy values. Specifically, four data sets called \emph{SubjectSet}, \emph{ResourceSet},
\emph{ActionSet} and \emph{EnvironmentSet} are defined.
These sets are filled with the values of the attributes of the policy elements \texttt{<Subjects>}, \texttt{<Resources>},
\texttt{<Actions>} and \texttt{<Environments>}, respectively.

Then, an ordered set of combinations of \emph{subject}, \emph{resource}, \emph{action} and
\emph{environment} is generated considering first Pairwise, then Threewise and finally Fourwise combinations.
Thus, the maximum number of requests derived by this strategy is equal to the cardinality of the Fourwise set.

The application of \emph{Simple Combinatorial} strategy to Policy 1 derives a set of 6 XACML requests, each one
containing one of the possible combinations of the subject, resource, action
and environment values taken from the policy.
An example of XACML request generated by the \emph{Simple Combinatorial} for Policy 1 is shown in Figure \ref{request-exampleSimple}. This request specifies that subject \emph{Julius} wants to \emph{write} on \emph{books} resource.

\subsubsection{Multiple Combinatorial}

Similarly to \emph{Simple Combinatorial}, the \emph{Multiple Combinatorial} strategy relies on a
combinatorial approach of the policy values. In particular, starting from the \emph{SubjectSet}, \emph{ResourceSet}, \emph{ActionSet}, \emph{EnvironmentSet} as defined above, this strategy derives for each set \emph{S}, the power set of S, called P(S), as the set of all possible subsets of S. The cardinality of P(S) = $2^{n}$, where \emph{n} is the
cardinality of S.
The test requests are then generated by combining the
subject, resource, action and environment subsets using Pairwise, Threewise, Fourwise combinatorial approaches.
The
maximum number of requests derived by this strategy is
equal to the cardinality of Fourwise set.

The application of \emph{Multiple Combinatorial} strategy to Policy 1 derives a set of 64 XACML requests, each one
containing one or more subject, resource, action
and environment values taken from the policy.

An example of XACML request generated by the \emph{Multiple Combinatorial} for Policy 1 is shown in Figure \ref{request-exampleMultiple}. This request specifies that subject \emph{Julius}, who is also a \emph{professor}, wants to \emph{write} on the \emph{books} resource. With respect to the request of Figure \ref{request-exampleSimple}, this one includes two subjects (\emph{Julius} and \emph{professor}) with different attribute-id (\emph{subject-id} and \emph{role}).

\subsection{Two-Round application example}
\label{tworound}

In this section we present a two-round application example useful for better highlight the infrastructure behavior and features.
The policy adopted in this experiment (called Policy 1 in this paper) is a simplified version of an XACML policy defining the access rights to a library (Figure \ref{policy-example}). It includes a policy set target (line 3) that is empty; a policy target (lines 8-15) allowing the access only to the \emph{books} resource; a first rule (\emph{ruleA})
(lines 19-36) with a target (lines 20-35) specifying that this rule applies only
to the access requests of a \emph{read} action of \emph{books} resource with any environment; a second rule (\emph{ruleB}) (lines 40-55), which effect  is \emph{Deny} when the subject is ``Julius'', the action is ``write'', the resource and environment are any resource and any environment respectively; a third rule (\emph{ruleC}) (lines 59-81) that allows to subject ``Julius'' the action ``write'', if he is also ``professor'' or ``administrator''; finally, the default rule (line 85) denies the access in the other cases.

\subsubsection{First round testing}
\label{firstround}

\begin{table*}[ht!]
\centering
\caption{\emph{Rule Target True/False Criteria} of Policy 1}
\label{table:traceRTTF}
\begin{tabular}{c}
\hline
A: Rule Target True Criterion \\
\hline
$T_{1}$ = \{($\emptyset$, $\emptyset$, $\emptyset$, $\emptyset$), ($\emptyset$, \{$books$\}, $\emptyset$, $\emptyset$),($\emptyset$, \{$books$\}, \{$read$\}, $\emptyset$), $Permit$\} \\
$T_{2}$ =  \{($\emptyset$, $\emptyset$, $\emptyset$, $\emptyset$), ($\emptyset$, \{$books$\}, $\emptyset$, $\emptyset$),(\{$Julius$\}, \{$books$\}, \{$write$\}, $\emptyset$), $Deny$\} \\
$T_{3}$ =  \{($\emptyset$, $\emptyset$, $\emptyset$, $\emptyset$), ($\emptyset$, \{$books$\}, $\emptyset$, $\emptyset$),(\{$Julius$\}, \{$books$\}, \{$write$\}, $\emptyset$), $-$\} \\
$T_{4}$ =  \{($\emptyset$, $\emptyset$, $\emptyset$, $\emptyset$), ($\emptyset$, \{$books$\}, $\emptyset$, $\emptyset$),($\emptyset$, $\emptyset$, $\emptyset$, $\emptyset$), $Deny$\} \\
\hline
B: Rule Target False Criterion \\
\hline
$T_{1}$ =  \{($\emptyset$, $\emptyset$, $\emptyset$, $\emptyset$), ($\emptyset$, \{$books$\}, $\emptyset$, $\emptyset$),($\emptyset$, \{$books$\}, \{$\neq read$\}, $\emptyset$), $-$\} \\

$T_{2_1}$ =  \{($\emptyset$, $\emptyset$, $\emptyset$, $\emptyset$), ($\emptyset$, \{$books$\}, $\emptyset$, $\emptyset$),(\{$ \neq Julius$\}, \{$books$\}, \{$\neq write$\}, $\emptyset$), $-$\} \\

$T_{2_2}$ =  \{($\emptyset$, $\emptyset$, $\emptyset$, $\emptyset$), ($\emptyset$, \{$books$\}, $\emptyset$, $\emptyset$),(\{$ \neq Julius$\}, \{$books$\}, \{$write$\}, $\emptyset$), $-$\} \\

$T_{2_3}$ =  \{($\emptyset$, $\emptyset$, $\emptyset$, $\emptyset$), ($\emptyset$, \{$books$\}, $\emptyset$, $\emptyset$),(\{$Julius$\}, \{$books$\}, \{$ \neq write$\}, $\emptyset$), $-$\} \\

$T_{3_1}$ =  \{($\emptyset$, $\emptyset$, $\emptyset$, $\emptyset$), ($\emptyset$, \{$books$\}, $\emptyset$, $\emptyset$),(\{$ \neq Julius$\}, \{$books$\}, \{$\neq write$\}, $\emptyset$), $-$\} \\

$T_{3_2}$ =  \{($\emptyset$, $\emptyset$, $\emptyset$, $\emptyset$), ($\emptyset$, \{$books$\}, $\emptyset$, $\emptyset$),(\{$ \neq Julius$\}, \{$books$\}, \{$write$\}, $\emptyset$), $-$\} \\

$T_{3_3}$ =  \{($\emptyset$, $\emptyset$, $\emptyset$, $\emptyset$), ($\emptyset$, \{$books$\}, $\emptyset$, $\emptyset$),(\{$Julius$\}, \{$books$\}, \{$ \neq write$\}, $\emptyset$), $-$\} \\

\hline

\end{tabular}
\end{table*}

\begin{table*}[]
\centering
\caption{\emph{Rule Condition True/False Criteria} of Policy 1}
\label{table:traceRCTF}
\begin{tabular}{c}
\hline
A: Rule Condition True Criterion \\
\hline
$T_{1}$ = \{($\emptyset$, $\emptyset$, $\emptyset$, $\emptyset$), ($\emptyset$, \{$books$\}, $\emptyset$, $\emptyset$),($\emptyset$, \{$books$\}, \{$read$\}, $\emptyset$), $Permit$\} \\
$T_{2}$ = \{($\emptyset$, $\emptyset$, $\emptyset$, $\emptyset$), ($\emptyset$, \{$books$\}, $\emptyset$, $\emptyset$),(\{$Julius$\}, \{$books$\}, \{$write$\}, $\emptyset$), $Deny$\} \\
%
$T_{3}$ = \{($\emptyset$, $\emptyset$, $\emptyset$, $\emptyset$), ($\emptyset$, \{$books$\}, $\emptyset$,
$\emptyset$),(\{$Julius$ $\land$ \{$professor$ $\lor$ $administrator$\}\}, \{$books$\}, \{$write$\}, $\emptyset$), $Permit$\} \\

$T_{4}$ =  \{($\emptyset$, $\emptyset$, $\emptyset$, $\emptyset$), ($\emptyset$, \{$books$\}, $\emptyset$, $\emptyset$),($\emptyset$, $\emptyset$, $\emptyset$, $\emptyset$), $Deny$\} \\
\hline
B: Rule Condition False Criterion \\
\hline
$T_{3}$ =  \{($\emptyset$, $\emptyset$, $\emptyset$, $\emptyset$), ($\emptyset$, \{$books$\}, $\emptyset$, $\emptyset$),(\{$Julius$ $\land$ \{$\neq professor \land \neq administrator$\}\}, \{$books$\}, \{$write$\}, $\emptyset$), $-$\} \\
\hline
\end{tabular}
\end{table*}

In the first phase of the experiment, we applied the \emph{Simple Combinatorial} strategy to  the policy of Figure \ref{policy-example}.
Each of the 6 generated XACML requests is transformed into an event by the \emph{Test case executor} component and  sent to the \emph{Monitor infrastructure}.

Starting from the policy, for each coverage criteria defined in Section \ref{coverageCriteria}, the \emph{Trace generator} derives the set of traces satisfying it.
In Tables \ref{table:traceRTTF} and \ref{table:traceRCTF}, the set of traces derived according to \emph{Rule Target True  Criterion}, \emph{Rule Target False Criterion}, \emph{Rule Condition True Criterion} and \emph{Rule Condition False Criterion} are provided.
Each \emph{Trace Set} is sent to the \emph{Monitoring Infrastructure} as a event.

At this point the  \emph{Policy evaluation engine} is instantiated with the XACML Sun' Policy Decision Point (PDP) \cite{pdp} which executes the XACML requests against the XACML policy and
sends the corresponding XACML responses to the \emph{Monitor infrastructure}.
The \emph{Monitor infrastructure} observes the on-line execution  of the XACML policy on the PDP, and, according to the values of the requests, the responses and the set of traces generated from the XACML policy, assesses the coverage of the XACML requests on the traces.

As an example, in Figure \ref{rule} there is a rule definition for trace $T_{3}$ of Table \ref{table:traceRCTF}A.
This rule is specified in Drools language \cite{droolsdoc} which is the reference language of the Glimpse monitor infrastructure used in this implementation. It is derived by the monitor infrastructure as in the following. First, the monitor infrastructure extracts from the payload of the event \emph{GlimpseBaseEventRequest} the field data that contains the values for the ($Subjects$, $Resources$, $Actions$, $Environments$) and then checks if they are included in the sets of $policySet$, $policy$ and $rules$ target values of the trace $T_{3}$ (lines [21-37]).
 If this is verified the monitoring infrastructure extracts from the same trace the $Response$ value and, if it is not empty, the monitor checks whether it is equal to the corresponding PDP response (line 43).  If this is true the trace is considered covered (line 49).

In this first round of the experiment, we executed the XACML policy of Figure \ref{policy-example} with all the requests generated by the \emph{Simple-Combinatorial} strategy and we reached the coverage results presented in Table \ref{coveragePolicy1}.

Specifically, each row of the table is associated to a different test criterion, while in the second column the results of the coverage obtained for the test cases derived by the \emph{Simple-Combinatorial} are reported. As in the table, all the criteria except  \emph{Rule Condition True Criterion} reached the 100\% coverage of the traces set.
In particular, for the \emph{Rule Condition True Criterion} only $T_{1}$, $T_{2}$, $T_{4}$ were covered.

Specifically, the request of Figure \ref{request-exampleSimple} is able to cover $T_{2}$ trace. Similarly, other requests having \emph{read} as action,  \emph{books} as resource and any value for subject are able to cover $T_{1}$ trace; whereas requests having any subject, action, and environment are able to cover $T_{4}$ trace.
From the analysis of the coverage assessment results, it was evident that, by construction, the test suite derived from \emph{Simple-Combinatorial} strategy  can only cover traces including only one subject, resource, action and environment value.
The coverage of trace $T_{3}$  requires XACML requests having more than
one subject values because the effect of the corresponding XACML rule (\emph{Rule C} of Figure \ref{policy-example}) is simultaneously dependent on more than one subject constraints.

This first round experiment evidenced two peculiarities: the former is that the \emph{Simple-Combinatorial} strategy is not effective enough to reach 100\% coverage of the traces and should be enriched.
By the identification of not covered traces, the Policy Assessment infrastructure  provides important hints to testers and can guide them in the generation of ad hoc test cases or selection of more effective test strategies. The latter concerns the policy specification itself. Specifically, the application of different coverage criteria, in particular the \emph{Rule Condition False Criterion} showed that the behavior of the system defined in \emph{Rule B} is the same as that defined in \emph{Rule C} when its condition is evaluated to False.
In particular, the behavior specified by Rule B is that \emph{Julius} cannot \emph{write} on the \emph{books} resource. This is the same expressed by Rule C when its  Condition  is evaluated to \emph{False}.
This implies a redundancy of the Policy 1.

Considering the results obtained by the \emph{Simple Combinatorial} test strategy, to improve the coverage in case of \emph{Rule Condition True} criterion, we adopted a different test generation approach. In particular, we used the proposed framework with  the XACML policy of Figure \ref{policy-example} and the requests generated by the \emph{Multiple-Combinatorial} strategy. As shown in Table \ref{coveragePolicy1}, third column, in this case we reached the 100\% coverage results for all the four criteria.


\begin{table}[ht]
\caption{Coverage Results for Policy 1}
\label{coveragePolicy1}
\centering
\resizebox{1.0\columnwidth}{!}{
\begin{tabular}{c c c }
\hline\hline
Coverage Criterion & Simple Combinatorial (6)  & Multiple Combinatorial (64) \\
\hline
Rule Target True & 100\% & 100\% \\
Rule Target False & 100\% & 100\% \\
Rule Condition True & 75\% & 100\% \\
Rule Condition False & 100\% & 100\% \\
\hline
\end{tabular}
}
\end{table}

\begin{table*}[h!]
\centering
\caption{\emph{Rule Target True/False Criteria} of Policy 2}
\label{table:traceRTTF2}
\begin{tabular}{c}
\hline
A: Rule Target True Criterion \\
\hline
$T_{1}$ = \{($\emptyset$, $\emptyset$, $\emptyset$, $\emptyset$), ($\emptyset$, \{$books$\}, $\emptyset$, $\emptyset$),($\emptyset$, \{$books$\}, \{$read$\}, $\emptyset$), $Permit$\} \\
$T_{2}$ =  \{($\emptyset$, $\emptyset$, $\emptyset$, $\emptyset$), ($\emptyset$, \{$books$\}, $\emptyset$, $\emptyset$),(\{$Julius$\}, \{$books$\}, \{$write$\}, $\emptyset$), $Deny$\} \\
$T_{3}$ =  \{($\emptyset$, $\emptyset$, $\emptyset$, $\emptyset$), ($\emptyset$, \{$books$\}, $\emptyset$, $\emptyset$),(\{$Marc$\}, \{$books$\}, \{$write$\}, $\emptyset$), $-$\} \\
$T_{4}$ =  \{($\emptyset$, $\emptyset$, $\emptyset$, $\emptyset$), ($\emptyset$, \{$books$\}, $\emptyset$, $\emptyset$),($\emptyset$, $\emptyset$, $\emptyset$, $\emptyset$), $Deny$\} \\
\hline
B: Rule Target False Criterion  \\
\hline
$T_{1}$ =  \{($\emptyset$, $\emptyset$, $\emptyset$, $\emptyset$), ($\emptyset$, \{$books$\}, $\emptyset$, $\emptyset$),($\emptyset$, \{$books$\}, \{$\neq read$\}, $\emptyset$), $-$\} \\

$T_{2_1}$ =  \{($\emptyset$, $\emptyset$, $\emptyset$, $\emptyset$), ($\emptyset$, \{$books$\}, $\emptyset$, $\emptyset$),(\{$ \neq Julius$\}, \{$books$\}, \{$\neq write$\}, $\emptyset$), $-$\} \\

$T_{2_2}$ =  \{($\emptyset$, $\emptyset$, $\emptyset$, $\emptyset$), ($\emptyset$, \{$books$\}, $\emptyset$, $\emptyset$),(\{$ \neq Julius$\}, \{$books$\}, \{$write$\}, $\emptyset$), $-$\} \\

$T_{2_3}$ =  \{($\emptyset$, $\emptyset$, $\emptyset$, $\emptyset$), ($\emptyset$, \{$books$\}, $\emptyset$, $\emptyset$),(\{$Julius$\}, \{$books$\}, \{$ \neq write$\}, $\emptyset$), $-$\} \\

$T_{3_1}$ =  \{($\emptyset$, $\emptyset$, $\emptyset$, $\emptyset$), ($\emptyset$, \{$books$\}, $\emptyset$, $\emptyset$),(\{$ \neq Marc$\}, \{$books$\}, \{$\neq write$\}, $\emptyset$), $-$\} \\

$T_{3_2}$ =  \{($\emptyset$, $\emptyset$, $\emptyset$, $\emptyset$), ($\emptyset$, \{$books$\}, $\emptyset$, $\emptyset$),(\{$ \neq Marc$\}, \{$books$\}, \{$write$\}, $\emptyset$), $-$\} \\

$T_{3_3}$ =  \{($\emptyset$, $\emptyset$, $\emptyset$, $\emptyset$), ($\emptyset$, \{$books$\}, $\emptyset$, $\emptyset$),(\{$Marc$\}, \{$books$\}, \{$ \neq write$\}, $\emptyset$), $-$\} \\
\hline
\end{tabular}
\end{table*}


\subsubsection{Second round testing}
\label{secondRound}

In the second round of the experiment, by the analysis  of the coverage results, we fixed the inconsistences detected in Policy 1 during the test case execution.
The resulting modified policy, named Policy 2 in the paper, is equal to the policy of Figure \ref{policy-example} except for line 63 in which the subject value is \emph{Marc} instead of \emph{Julius}.
Starting from Policy 2, the \emph{Trace Generator} derived the sets of traces of Tables \ref{table:traceRTTF2} and \ref{table:traceRCTF2}, according to \emph{Rule Target True Criterion}, \emph{Rule Target False Criterion}, \emph{Rule Condition True Criterion} and \emph{Rule Condition False Criterion} of Section \ref{coverageCriteria}.

As in the first round experiment, we used combinatorial strategies for generating the sets of test cases from Policy 2. Specifically, we derived 8 and 128 test cases according to \emph{Simple Combinatorial} and \emph{Multiple Combinatorial} strategies respectively.
We used the proposed framework for executing these test cases against all the four criterion defined in Section \ref{coverageCriteria}. The obtained coverage results are presented in Table \ref{coveragePolicy2} (second and third column).

\begin{table*}[]
\centering
\caption{\emph{Rule Condition True/False Criteria} of Policy 2}
\label{table:traceRCTF2}
\begin{tabular}{c}
\hline
A: Rule Condition True Criterion \\
\hline
$T_{1}$ = \{($\emptyset$, $\emptyset$, $\emptyset$, $\emptyset$), ($\emptyset$, \{$books$\}, $\emptyset$, $\emptyset$),($\emptyset$, \{$books$\}, \{$read$\}, $\emptyset$), $Permit$\} \\
$T_{2}$ =  \{($\emptyset$, $\emptyset$, $\emptyset$, $\emptyset$), ($\emptyset$, \{$books$\}, $\emptyset$, $\emptyset$),(\{$Julius$\}, \{$books$\}, \{$write$\}, $\emptyset$), $Deny$\} \\
%
$T_{3 }$ =  \{($\emptyset$, $\emptyset$, $\emptyset$, $\emptyset$), ($\emptyset$, \{$books$\}, $\emptyset$,
$\emptyset$),(\{$Marc$ $\land$ \{$ professor$ $\lor$ $administrator$\} \}, \{$books$\}, \{$write$\}, $\emptyset$), $Permit$\} \\

$T_{4}$ =  \{($\emptyset$, $\emptyset$, $\emptyset$, $\emptyset$), ($\emptyset$, \{$books$\}, $\emptyset$, $\emptyset$),($\emptyset$, $\emptyset$, $\emptyset$, $\emptyset$), $Deny$\} \\
\hline
B: Rule Condition False Criterion \\
\hline
$T_{3}$ =  \{($\emptyset$, $\emptyset$, $\emptyset$, $\emptyset$), ($\emptyset$, \{$books$\}, $\emptyset$, $\emptyset$),(\{$Marc$ $\land$ \{$\neq professor$ $\land$ $\neq administrator$\}\}, \{$books$\}, \{$write$\}, $\emptyset$), $-$\} \\
\hline
\end{tabular}
\end{table*}

The coverage results of this second round experiment confirm those of the first round. As before the
\emph{Simple Combinatorial} strategy does not
reach the 100\% coverage whereas \emph{Multiple Combinatorial} does it.

Moreover, due to the higher number of subjects in Policy 2, the number of derived requests is higher than that of Policy 1 both for \emph{Simple Combinatorial} and \emph{Multiple Combinatorial} strategies.

\begin{table}[ht]
\caption{Coverage Results for Policy 2}
\label{coveragePolicy2}
\centering
\resizebox{1.0\columnwidth}{!}{
\begin{tabular}{c c c }
\hline\hline
Coverage Criterion & Simple Combinatorial (8)  & Multiple Combinatorial (128) \\
\hline
Rule Target True & 100\% & 100\% \\
Rule Target False & 100\% & 100\% \\
Rule Condition True & 75\% & 100\% \\
Rule Condition False & 100\% & 100\% \\
\hline
\end{tabular}
}
\end{table}

\subsection{Experimental results}
\label{experiment}
In this section, we show the application of the Policy Assessment Infrastructure to three real-world XACML policies. In particular,   we used the  policies (named demo-5, demo-11 and demo-26) taken from the Open Source repository software Fedora (Flexible Extensible
Digital Object Repository Architecture) \cite{fedora} for controlling the access to the
administered digital contents. Details about the policies structure are reported in Table \ref{tab:PoliciesSubjects}.  In particular, the columns represent the number of rules, conditions, subjects, resources, actions and distinct functions within each
policy.

The coverage experimental results are reported in Table \ref{experimental}. In particular,  for each of the considered policies the rows  reports, the applied  test criterion (first column), the number of relative traces (second column), the number of the requests generated by the \emph{Simple-Combinatorial} and \emph{Multiple-Combinatorial} test strategies (columns 3 and 5 respectively) and the  coverage results obtained by each test suite (Columns 4 and 6 respectively).

As for the  two-round application example, also in this experiment the \emph{Multiple-Combinatorial} test strategy always outperforms the  \emph{Simple-Combinatorial} one. This is again due to the possibility provided by the \emph{Multiple-Combinatorial} of generating  requests having multiple elements.

This is particulary evident considering the   \emph{Rule Condition False Criterion} where the \emph{Simple-Combinatorial} reached the 33,50\%, 0\%, 14,28\% coverage percentage of the traces set for demo5, demo26, demo11 respectively.
Indeed, according to Definition \ref{def:rcf}, to satisfy such criterion,  it is first necessary  to satisfy the \emph{Rule Target True} criterion and make the  condition evaluated to false. For all the policies, the target of the rule considered contains more than one resource value, therefore cannot be never satisfied by the test cases generated by \emph{Simple-Combinatorial} test strategy.

However, considering the cardinality of the two test suites the   \emph{Multiple-Combinatorial} is much more greater than  the  \emph{Simple-Combinatorial} one. This number could reach not manageable values, as for instance for \emph{demo5}, invalidating the efficacy of the \emph{Multiple-Combinatorial} test strategy.
In these cases, the analysis of the coverage percentage of the different criteria could be used for increasing the performance of simpler test strategies (as for instance \emph{Simple Combinatorial}), by adding just the ad hoc test cases so to reach an established coverage boundary.

These experiment results confirm the ones reported in the previous section and highlight better how  the analysis of the coverage percentage of the different criteria could be a practical means for evaluating and improving the test strategies.

\begin{table}[ht]
\caption{XACML Policies Subjects.} \label{tab:PoliciesSubjects}
\begin{center}
\begin{tabular}{lllllll}
\hline
Xaml Policy & \multicolumn{6}{c}{Functionality}\\
\hline
& \# Rule & \# Cond & \# Sub & \# Res & \# Act & \# Funct \\
\hline
demo-5 & 3  & 2  & 6  & 3  & 2  & 4 \\
demo-11 & 3  & 2  & 4  & 3  & 1  & 5 \\
demo-26 & 2  & 1  & 1  & 3  & 1  & 4 \\
\hline
\end{tabular}
\end{center}
\end{table}

\begin{table*}[ht]
\caption{Experimental Results}
\label{experimental}
\centering
\begin{tabular}{|c|c|c|c|c|c|}
\hline
Coverage Criterion &  Traces ($\sharp$) & Requests ($\sharp$) & Coverage (\%) & Requests ($\sharp$) & Coverage (\%) \\

\hline
\multicolumn{6}{|c|}{demo5} \\
\hline
 & & \multicolumn{2}{|c|}{Simple Combinatorial} & \multicolumn{2}{|c|}{Multiple Combinatorial} \\
\hline
Rule Target True & 4 & 36 & 50\% & 2048 & 100\% \\
\hline
Rule Target False & 7 & 36 & 71,50\% & 2048 & 100\% \\
\hline
Rule Condition True & 7 & 36 & 50\% & 2048 & 100\% \\
\hline
Rule Condition False & 9 & 36 & 33,50\% & 2048 & 100\% \\
\hline
\multicolumn{6}{|c|}{demo11} \\
\hline
 & & \multicolumn{2}{|c|}{Simple Combinatorial} & \multicolumn{2}{|c|}{Multiple Combinatorial} \\
 \hline
Rule Target True & 4 & 12 & 50\% & 256 & 100\% \\
\hline
Rule Target False & 6 & 12 & 66,67\% & 256 & 100\% \\
\hline
Rule Condition True & 4 & 12 & 50\% & 256 & 100\% \\
\hline
Rule Condition False & 7 & 12 & 14.28\% & 256 & 100\% \\
\hline
\multicolumn{6}{|c|}{demo26} \\
\hline
 & & \multicolumn{2}{|c|}{Simple Combinatorial} & \multicolumn{2}{|c|}{Multiple Combinatorial} \\
\hline
Rule Target True & 3 & 3 & 33\% & 32 & 100\% \\
\hline
Rule Target False & 6 & 3 & 66,67\% & 32 & 100\% \\
\hline
Rule Condition True & 3 & 3 & 33\% & 32 & 100\% \\
\hline
Rule Condition False & 2 & 3 & 0\% & 32 & 100\% \\
\hline
\end{tabular}
\end{table*}

\section{Related Work}
\label{related}

In software testing, white-box
approaches based on the coverage of specified entities are
considered a valuable complement to black-box ones,
as coverage information can provide an indication of the
thoroughness of the executed test cases, and can help to
maintain an effective test suite.
Code coverage criteria are
also addressed by most of the techniques of test cases
prioritization \cite{kaur2011genetic,leon2003comparison}, with the aim to reorder test cases so that those tests that have a higher priority are
executed before the ones having a lower priority.
Many proposals for test coverage measurement and analysis have been
proposed depending on the adopted policy specification language\cite{shahid2011study} .

 In literature there are few works facing coverage assessment of XACML policies. Seminal works are presented in  \cite{martin2006defining} and \cite{bertolino2014coverage}. In the former, the authors provide a first coverage criterion for XACML policies defining three
structural coverage metrics targeting XACML policies, rules
and conditions respectively. These coverage
metrics are used for reducing test sets and the effects of
test reduction in terms of fault detection are measured. In the latter, the authors also address the policy set and do not require the policy execution and PDP instrumentation. The focus is in  reducing the effort for coverage measurement focusing explicitly on the evaluation of the  Rule Target Set.
More recently, in \cite{8003052} the authors extend the introduced concept by presenting a combinatorial testing approach based on data flow coverage, while in \cite{Xu:2016} the test execution information is used to determine which policy element is faulty. Specifically, the authors introduced the concept of reachability-based and firing-based fault localization techniques.
In this paper, we extend the criteria proposed in \cite{bertolino2014coverage} differentiating the concept of  Rule
Target matching, so to better analyze the evaluation of policy rules.

Considering the automated test cases generation, solutions have been proposed for testing either the XACML policy or the PDP implementation \cite{daoudagh2015assessment,iet,bertolino2014testing}.
Among them, the most referred ones are the Targen tool~\cite{martin06:automated}, the proposal of \cite{pretschner2008model} and the already mentioned X-CREATE tool \cite{bertolino2012automatic,seaa10, iet}.

Finally, concerning the adoption of monitoring facilities for coverage assessment, the work in \cite{telerise17} represents a preliminary attempt of an infrastructure for enabling coverage criterion selection and policy coverage assessment analysis.
Several general-purpose monitoring proposals are currently available, which can be mainly divided into
two groups: those that are embedded in the execution engine such as \cite{daoudagh2015assessment,mouelhi2015chapter} and those that can be integrated into the execution framework as an additional component such for instance \cite{carvallo2017multi}.
Among the additional monitor facility in this paper  we refer to the
monitoring framework called Glimpse \cite{bertolino2011towards}, which is extremely
flexible and adaptable to various scenarios and SOA architecture
patterns.

\section{Discussion and Conclusions}
\label{conclusion}

In this paper, we presented a set of coverage criteria and a policy assessment infrastructure. This last enables the coverage criterion selection, the monitoring of the policy execution and the analysis of the policy coverage assessment. The conceived architecture would like to be independent from any specification language. Moreover, we provided an instantiation inside the XACML-based access control systems and its evaluation on real XACML policies.

The collected results, even if preliminary, showed from one side the effectiveness of the proposed infrastructure in evaluating the coverage of an XACML policy, from the other evidenced some inconsistencies in the policy specification. The analysis of the coverage assessment highlighted weaknesses in the test suite and provided hints both for the generation of ad-hoc test cases and suitable modification of the policy specification.

Concerning the computational performance of the policy assessment structure, because the framework requires just to listen the messages flow and works independently from the access control system considered,  it does not impact on the overall performance on this last. Moreover, as described in Section \ref{framework}, the compositional nature of the four main components, lets a customizable implementation of different test strategies and testing criteria. Indeed,  tools and solutions can be easily integrated provided that the specific interaction message formats are guaranteed.
Considering in detail the implementation of the policy assessment structure used in this paper, the \emph{Test generator} and the \emph{Trace generator} components have only
 off-line activities and do not impact on the online computational performance of the proposed framework. The most critical component for the on-line testing activity is the \emph{Monitor} one. In this implementation we used the Glimpse infrastructure, which in turn relies on Drools RETE matching algorithms computational performance. We refer to \cite{droolsdoc} for more details.

In this paper, the usage and the experimentation of the proposed access control infrastructure has been limited to the coverage policy assessment, however it can have an important role also  to tracking the actual resources access and users behavior.

Concerning threats to validity of the presented experiment,
the tools adopted and the algorithms implemented may have influenced the reported results. It could be that different choices might have provided different effectiveness results.
Another threat to our proposal is due to the employed test strategies.
We used Simple and Multiple Combinatorial test strategies but it is likely that other test
sets may produce different coverage results.
Finally, external validity of the experiment concerns potential issues
that may prevent the generalization of the results, as in all empirical studies including ours.

We are currently working to include in the access control infrastructure more coverage criteria. We plan also to enhance the monitor infrastructure with facilities for proactively detecting, by the off-line trace analysis,  possible security inconsistencies of the tested access control policy. Other future work deals with the instantiation of the proposed infrastructure by considering different access and usage control policy specification languages.
Finally, part of future work would be also to compare our approach with others XACML based coverage assessment solutions based on the instrumentation of the PDP engine.

\section*{Acknowledgments}

This work has been partially supported by the GAUSS national research
project (MIUR, PRIN 2015, Contract 2015KWREMX).
We also thank Antonello Calabr\`{o} for his help during the integration of Glimpse infrastructure.

\newpage

\bibliographystyle{acm}

\bibliography{biblio}

\end{document}